# Derivation of Tsallis statistics from dynamical equations for a granular gas

**F. Sattin**

*Consorzio RFX, Associazione Euratom-ENEA per la Fusione,*

*Corso Stati Uniti 4, 35127 Padova - Italy*

*e-mail: sattin@igi.pd.cnr.it*

**Abstract**

In this work we present the explicit calculation of velocity Probability Distribution Function $P(v)$ for a model system of granular gas within the framework of Tsallis Non-Extensive Statistical Mechanics, using the stochastic approach by Beck [C. Beck, Phys. Rev. Lett. **87**, 180601 (2001)], further generalized by Sattin and Salasnich [F. Sattin and L. Salasnich, Phys. Rev. E **65**, 035106(R) (2002)]. The calculation is self-consistent in that the form of $P(v)$ is not given as an ansatz but is shown to necessarily arise from the known microscopic dynamics of the system.

# I. Introduction

A large amount of research has recently been devoted to the study of spatially uniform systems with dissipative interactions ("granular matter") (see, e.g., [1-6] and references therein). Among the issues addressed, much attention is deserved to transport properties and therefore to the structure of the asymptotic velocity statistics. Although details vary according to the particular model chosen for describing the granular matter, an interesting feature common to almost all of these models is the overpopulation of the high-velocity tail of the Probability Distribution Function (PDF) with respect to the standard Maxwellian (Gaussian) distribution, either in the form of a stretched exponential, or a power-law distribution [5].

The existence of non-Gaussian high-energy tails calls obviously for points of contact with the formalism of the Nonextensive Statistical Mechanics (NESM), originally introduced by Tsallis [7], which has exactly this distinctive feature. Furthermore, there is the interesting property of some models for granular materials (so-called Inelastic Maxwell Models-IMM) of allowing performing analytical calculations. The fact that velocity PDF's can be computed exactly from the constitutive equations of the model raises the interesting question: how Tsallis statistics is actually generated by the microscopic dynamics of a system. This point, in connection with granular gases, was first raised but left unanswered in [1]. In this paper we will address this question and show that PDF's for the class of granular gases studied can be computed self-consistently by essentially combining the microscopic dynamics with the hypothesys of the molecular chaos and with a stochastic formulation of NESM. To the best of our knowledge, only another group [8] has been able till now to explicitly derive Tsallis statistics from dynamical equations of an Hamiltonian system (but see also Appendix A). However, there, the system was built *ad hoc*. In our case, instead, we are studying a model originally devised for a completely different purpose and which aims to be an approximate but realistic approximation of physical systems found in reality.

The structure of this paper is as follows: in section II we will give a brief description of the IMM; section III will be devoted to the description of the stochastic interpretation of NESM. Calculations will be performed in section IV. Finally, section V will be devoted to conclusions.

## II. The Inelastic Maxwell Model

The model we use, or some variants of it, has been extensively used for the study of one- or two-component granular materials [1-6]. Granular materials are assemblies of macroscopic particles interacting through inelastic collisions. They possess some uncommon features; notably, fail to obey the zeroth law of thermodynamics: in a mixture of two or more interacting components isolated from the outside, each component can reach a temperature (measured as the average kinetic energy of its elements) different from the other components, although in presence of an exchange of energy among them [1, 2].

Let us consider an ensemble of $N$ particles, $N_1$ of species 1 and $N_2$ of species 2. Let $p = N_1/(N_1 + N_2) = N_1/N$. The masses of the two species are different, respectively $m_1$ and $m_2$ and we define $\zeta = m_1/m_2$. To fix ideas, also, let $m_1 \geq m_2$ ($\zeta \geq 1$). The particles collide between them at a rate independent of their velocity and only binary impulsive collisions are considered. Collisions can be inelastic and the degree of loss of energy in a collision between two particles of species $\alpha$ and $\beta$ is quantified by the restitution coefficient $r_{\alpha\beta} \leq 1$, with the equal sign holding for elastic collisions.

In one-dimensional geometry, the velocities of two particles $i$ and $j$ change after each collision according to

$$v_i^{'(\alpha)} = v_i^{(\alpha)} - (1 + r_{\alpha\beta}) \frac{m_\beta}{m_\alpha + m_\beta} (v_i^{(\alpha)} - v_j^{(\beta)})$$

$$v_j^{'(\beta)} = v_j^{(\beta)} + (1 + r_{\alpha\beta}) \frac{m_\alpha}{m_\alpha + m_\beta} (v_i^{(\alpha)} - v_j^{(\beta)})$$

(1)

The upperscript is to remind that the collision can be between like ($\alpha = \beta$) or unlike ($\alpha \neq \beta$) particles. It is possible to show that equations (1) conserve the momentum but not the energy, unless all the $r_{\alpha\beta}$'s are equal to 1. All the particles are allowed to interact, regardless of their position (mean-field approximation). In numerical calculations, a predefined fraction of particles at each time step is randomly picked up and made to collide.

Let us now briefly describe the main features of this model. Master equations can be derived for velocity PDF's $P_\alpha(v,t)$, which give the probability density for finding a particle of species α at time $t$ with a velocity around $v$. They satisfy a coupled set of differential equations (see [1,4]).

In the case of a pure component, i.e. only particles of species 1 or 2, it is possible to show [1] that an asymptotic solution (i.e., valid for $t \to \infty$) of the corresponding equation for $P(v,t)$ exists and has the shape:

$$P_s(v,t) = \frac{2}{\pi} \frac{1}{v_0(t)} \frac{1}{\left[ 1 + \left( \frac{v}{v_0(t)} \right)^2 \right]^2} \qquad , \qquad (2)$$

where $v_0(t)$ is an exponentially decaying function: $v_0(t) = v_0(0) \, exp(-\lambda(r) \, t)$, and $\lambda$ is a function of the restitution coefficient with $\lambda = 0$ if $r = 1$.

The result (2) is intriguing since it predicts a power-law tail for high $v$. The $P_s$ has the form commonly found within the framework of NESM but it derives from "exact" equations of motion, and not from variational principles involving the entropy and the energy of the system.

The two-component case is more complicated and equations can be solved analytically only to a certain extent and under some approximations. The $p \to 0$ limit was addressed by Ben-Naim and Krapivsky [4]. This limit allows having one population (the background fluid) freely evolving while the other (the impurity) is driven by the former. For $\zeta >> 1$ (i.e., light impurities embedded into a fluid of heavy particles), Ben-Naim and Krapivsky demonstrated that the PDF's for the impurities as well as for the fluid can generically be written as linear combinations of powers of Lorentzian functions:

$$P_{fluid,imp}(v) = \sum_{n \geq 2} B_{fluid,imp}^{(n)} \left( \frac{1}{1+v^2} \right)^n \qquad . \qquad (3)$$

The coefficients $B^{(n)}$ depend upon the restitution coefficients, and some or even infinitely many of them (but, of course, not all) may be zero. In the limit of high $v$, only the leading term survives, hence $P \sim 1/v^4$ but for particular combinations of the parameters. Of course, the result (3) encompasses (2) since $p = 0$ corresponds to the pure case.

It is worthwhile mentioning that Bettolo Marconi *et al*. [1] also addressed the two-component problem, still assuming $\zeta >> 1$, but allowing for finite values of $p$. They found numerically that both species of particles have leading terms for high-$v$ tails of the form

$$P_{\alpha,\beta}(v,t) \approx \frac{1}{\left[1 + \left(\frac{v}{v_0(t)}\right)^2\right]^{\varepsilon(\alpha,\beta)}} \qquad , \qquad (4)$$

with exponents $\varepsilon$ which depend from all the parameters $p$, $r_{\alpha\beta}$, $\zeta$ in a complicated fashion. However, while still $\varepsilon(1) = 2$, the light component was found to have a steeper descent: $\varepsilon(2) \sim 3$. We argue that the inconsistency with the results [4] is only apparent, and the result (4) is a particular case of (3), corresponding to a peculiar choice of parameters such that $B^{(2)} \sim 0$. The calculation will be put forth in the Appendix B.

Before ending this section, we will do some elaboration on Eqns. (1) that will be useful for the following. By defining $v_i^{(\alpha)} - v_j^{(\beta)} = v_{rel}$, $(1 + r_{\alpha\beta})\frac{m_\beta}{m_\alpha + m_\beta} = f_{\alpha\beta}$ , we reorder the former of (1):

$$\Delta v_i^{(\alpha)} = -f_{\alpha\beta}\, v_{rel} \qquad . \qquad (5)$$

From now on we will consider formally the r.h.s. of (5) as a known function. By averaging over several collisions it is clear that Eq. (5) describes a Brownian motion in the velocity space, of step $|f_{\alpha\beta}\, v_{rel}|$. It is convenient to turn Eq. (5) into the form of a differential equation:

$$\frac{dv^{(\alpha)}}{dt} = |\, f_{\alpha\beta} v_{rel}\, |\, L(t) \qquad , \qquad (6)$$

with $L(t)$ white noise of zero mean and whose variance may be chosen unity after a suitable choice of time scale. The minus sign appearing in (5) has been absorbed into the definition of $L$.

At this point it is convenient to formally turn the system (6) into a driven one by adding a term $F = -\gamma\, v^{(\alpha)}$. This turns Eq. (6) into the Langevin equation

$$\frac{dv^{(\alpha)}}{dt} = -\gamma v^{(\alpha)} + |\, f_{\alpha\beta} v_{rel}\, |\, L(t) \qquad , \qquad (7)$$

from which the asymptotic temperature can be computed:

$$\frac{1}{\beta} = \frac{f_{\alpha\beta}^2 v_{rel}^2}{\gamma} \qquad . \qquad (8)$$

A nonzero value of $\gamma$ corresponds to providing energy to (or draining from) the system by the outside. Since in a dissipative system $v_{rel}$ approaches asymptotically 0, we can let $\gamma \rightarrow$ 0 while simultaneously maintaining finite the ratio (8). In this way, β is defined but for a multiplicative constant, which is of no relevance to our purposes.

## III. The stochastic interpretation of NESM

Tsallis NESM has gained a considerable interest in these years because of its capability to describe a wealth of disparate phenomena (from anomalous diffusion, to turbulent systems, etc.) apparently unexplainable within the framework of the standard statistical mechanics. A constantly updated list of references is found in [9]; some recent reviews on the subject are [10].

A subject relatively less explored concerns the relationship between NESM and the microscopic dynamical properties of a given system: one would like being able to compute the PDF for one system given its evolution equations; or, at least, uniquely relate the features of the PDF (say, the non-extensive exponent $q$) to some microscopical property characterizing the system. We mention here the paper by Wilk and Wlodarczyk [11], which first related the non-extensive exponent $q$ to a definite microscopic property: in their case, the fluctuations of the cross section for stopping of high-energy cosmic rays. Beck [12] provided later a theoretical framework to their finding; Almeida [13], by a different path reached a similar conclusion. More recently, Sattin and Salasnich [14] generalized Beck's approach. A deterministic connection between NESM and microdynamics was provided in a recent work (Adib et al. [8]), showing how power-law PDF's arise in mesoscopic systems evolving under suitable homogeneous Hamiltonians: an explicit calculation of a power-law PDF was performed using a Hamiltonian of the Fermi-Pasta-Ulam type.

It is useful to briefly summarize the results [12,14], which will be used in this work: according to Beck, any system $S$ endowed with Hamiltonian $H$ in equilibrium with a surrounding heat reservoir $B$ follows the standard Boltzmann statistics, thus its canonical distribution is given by the Maxwell-Gibbs exponent: $\rho(H) \propto \exp(-\beta H)$. However, the reservoir itself may be not a simple system in equilibrium, but instead consisting of several subsystems interacting between them. The generalized temperature $1/\beta$, which quantifies the interaction between $S$ and $B$ might be therefore not a constant but a function varying in time, to all extent a stochastic variable characterized by its own PDF $P(\beta)$. Therefore, any attempt of determining $\rho(H)$ would yield only its average over $P(\beta)$:

$$\rho(H) = \int d\beta\, C_H(\beta) \exp(-\beta H) P(\beta) \qquad . \qquad (9)$$

$C_H(\beta)$ is the normalization constant for $\rho(H)$, which depends in general both on $\beta$ as well as on the particular form for $H$. Beck argues that $P(\beta)$ should belong to a particular class of functions ($\chi^2$-distributions): it is straightforward to show that, using the standard free-particle Hamiltonian $H = u^2/2$, one recovers the power-law behaviour for $\rho(u)$ first predicted by Tsallis on the basis of the maximum entropy principle. If, instead, $P(\beta)$ is a Dirac delta: $P(\beta) = \delta(\beta - \beta_0)$, i.e. when fluctuations of $\beta$ around its average value are small, the standard extensive statistics is recovered. The two limits are not exclusive since a $\chi^2$-function with infinite degrees of freedom approaches a Dirac delta.

Beck's is an ansatz about $P(\beta)$. Sattin and Salasnich showed that if, instead, this constraint is relaxed and $\beta$ is supposed to be a derived quantity, $\beta = \beta(c_1, c_2, ...)$, where the $c_i$'s are the actual fluctuating parameters that govern the dynamics of the systems, it is possible to widen the class of distributions $P(\beta)$ may belong to. As a consequence, different, even non-power-law functional forms for $\rho(u)$ were obtained. It was shown in [14] that some experimental distributions could be fitted using these generalized functional forms.

## IV. Calculation of velocity statistics

Our purpose in this paper is to recover, within the interpretation for NESM described in section III, the results of [4], namely that: I) the PDF for a pure granular gas has a power-like high-$v$ tail; II) the PDF for a light impurity in a background fluid has the same shape than the fluid's. We point out from the outset that our treatment will fall short of: I) delivering the value for the exponent of the power-law; II) giving account of the particular cases found in [1] where the impurity and the fluid have different exponents. We believe-and try to demonstrate in the Appendix B-that this latter shortcoming is a minor one since these cases should be rather exceptional, i.e., of null measure over the parameter space.

We will begin with the pure case. The starting point is Eq. (9); hence, one must know $P(\beta)$ in order to estimate the velocity PDF $P(v)$. At this stage, two ways are possible: one, is to accept Beck's ansatz and postulate that $P(\beta)$ is a $\chi^2$-distribution:

$$P(\beta) = \frac{1}{\Gamma\left(\frac{n}{2}\right)} \left[\frac{n}{2\beta_0}\right]^{n/2} \beta^{n/2-1} \exp\left(-\frac{n\beta}{2\beta_0}\right) \qquad . \qquad (10)$$

In (10) $n$ is the number of degrees of freedom and $\beta_0$ the average value of $\beta$.

If we replace expression (10) in (9), and for $H$ take the free-particle Hamiltonian: $H = v^2/2$, we can write

$$P(v) = K \int \beta^{1/2} \exp(-\beta v^2/2) P(\beta) d\beta \qquad , \qquad (11)$$

where the term $\beta^{1/2}$ stems from the normalization of $\rho(H)$, and all other constants terms have been collected into $K$. It is straightforward to see that the result of the integration is

$$P(v) = K \frac{1}{\left[1 + \left(\frac{v}{v_0}\right)^2\right]^{\frac{n+1}{2}}} \qquad , \qquad (12)$$

hence the result (3) is obtained for $(n+1)/2 = 2 \rightarrow n = 3$.

This path is not completely satisfactory since it leaves us with at least two unanswered questions: I) why $P(\beta)$ has the form (10); and II) why $n$ takes exactly that value.

The other approach is patterned after that of Ref. [14]. Rather than postulating a form for $P(\beta)$, we attempt to derive one from what we know of the system. This way, we will be able to give an answer to question (I), although not to (II).

In Eq. (8) a relationship was written between $\beta$ and the relative impact velocity. Now we recall the relation between $v_{rel}$ and the particle velocity and point out that the latter is actually a stochastic variable with its own PDF. Therefore, $v_{rel}$ and $\beta$ too must be considered as stochastic variables. $\beta$ in particular, inherits its PDF from that of $v_{rel}$: if $P(v_{rel})$, $P(\beta)$ are the PDF's for $v_{rel}$ and $\beta$, the simple relationship exists

$$P(\beta) d\beta = P(v_{rel}(\beta)) \left|\frac{dv_{rel}}{d\beta}\right| d\beta \sim P(v_{rel}(\beta)) \frac{1}{\beta^{3/2}} d\beta \qquad . \qquad (13)$$

Since $P(v_{rel})$ is related to $P(v)$, $P(\beta)$ is some-still unknown-function of $P(v)$.

Eq. (11) with $P(\beta)$ given by Eq. (13) is actually an implicit equation for $P(v)$. Although the integral can be removed by a Fourier transform, we will show that there is not a real advantage, since $P(\beta)$ is still an integral function of $P(v)$. Hence, we will adopt the more nâive procedure to guess an initial functional form for $P(v)$, use it for computing $P(v_{rel})$, $P(\beta)$ and hence, through (11), recover $P(v)$ again: the correct PDF must be a fixed point of this sequence of transformations.

As a starting guess, we will adopt obviously the expression (4): the joint probability for two particles $i$ and $j$ (assumed uncorrelated) to have the velocities $v_i$, $v_j$ is therefore

$$P(v_i)P(v_j)dv_i\,dv_j = K\frac{1}{\left[1+\left(\dfrac{v_i}{v_i^{(0)}}\right)^2\right]^{\sigma}}\frac{1}{\left[1+\left(\dfrac{v_j}{v_j^{(0)}}\right)^2\right]^{\sigma}}dv_i\,dv_j \qquad . \qquad (14)$$

Since the two particles are of the same species they have the same PDF, hence we have chosen the same exponent $\sigma$ for both. Also, the average velocity must be the same, thus $v_i^{(0)} = v_j^{(0)} \equiv 1$ after suitable rescaling. By converting to the variables $v_{rel} = v_i - v_j$, $\quad y = v_j$ we can get the marginal probability $P(v_{rel})$ :

$$P(v_{rel})dv_{rel} = K\left(\int \frac{1}{\left[1+\left(v_{rel}+y\right)^2\right]^{\sigma}}\frac{1}{\left[1+y^2\right]^{\sigma}}dy\right)dv_{rel} \qquad . \qquad (15)$$

From here on all irrelevant constant terms must be considered as automatically merged into the constant $K$, without the need of explicitly redefining it each time. The integral within parentheses can be analitically performed. However, let us recall that we are interested to the high-energy tail of $P(v)$: therefore only the asymptotic behaviour for $\beta \rightarrow 0$ and thus for $v_{rel} \rightarrow \infty$ is of interest, and it can easily be shown that

$$P(v_{rel})dv_{rel} \xrightarrow{v_{rel}\rightarrow\infty} \frac{K}{v_{rel}^{2\sigma}}dv_{rel} \qquad . \qquad (16)$$

From (8, 13) we get

$$\lim_{\beta\rightarrow 0} P(\beta) = K\beta^{\sigma-3/2} \qquad (17)$$

and, replacing into (11),

$$P(v) = K\int \beta^{\sigma-1}\exp\left(-\beta v^2/2\right)d\beta \sim \frac{1}{(v^2)^{\sigma}} \qquad . \qquad (18)$$

We are thus back to the initial expression (14). This is not a rigorous demonstration, since we have not shown that there is only one solution. However, the several constraints that must be satisfied by any physically realizable solution (it must be continuous with at least some of its derivatives, normalizable, monotonously decreasing for high $v$) severely restrict the class of potential candidates. It seems therefore enough to show that some other apparently plausible choices fail instead to satisfy the above equations. We will repeat thus calculations for the most obvious choice: the Gaussian PDF.

We can straightforwardly write the equivalent of Eq. (15):

$$P(v_{rel})dv_{rel} = K\left[\int \exp\left(-\left(v_{rel}+y\right)^2 - y^2\right)dy\right]dv_{rel}$$
$$= K\exp\left(-v_{rel}^2/2\right)dv_{rel}$$
$$\rightarrow P(\beta) = K\frac{1}{\beta^{3/2}}\exp\left(-\frac{\beta_0}{\beta}\right)$$

(19)

with $\beta_0$ constant term. Hence,

$$P(v) = \int \beta^{-1}\exp\left(-\beta v^2/2 - \beta_0/\beta\right)d\beta \approx K_0\left(\sqrt{2\beta_0 v^2}\right)$$
$$\sim \frac{1}{v}e^{-\sqrt{2\beta_0}\,v} \quad (v\rightarrow\infty)$$

(20)

with $K_0$ modified Bessel function of order 0. Since we do not recover the initial PDF, this choice must be ruled out.

The two-component case is not dealt with with substantial differences. We limit to vanishingly fractions of impurities, $p\sim0$. This means that, in Eq. (1), an impurity particle is likely to collide only with a fluid particle. The relative velocity becomes therefore $v_{rel} = v_{imp} - v_{fluid}$. We allow for the two populations to have similar PDF's but with different exponents, thus Eq. (14) becomes

$$P(v_{imp})P(v_{fluid})dv_{imp}dv_{fluid} = K\frac{1}{\left[1+\left(\frac{v_{imp}}{v_{imp}^{(0)}}\right)^2\right]^{\sigma_{imp}}}\frac{1}{\left[1+\left(\frac{v_{fluid}}{v_{fluid}^{(0)}}\right)^2\right]^{\sigma_{fluid}}}dv_{imp}dv_{fluid}$$

(21)

Now the two average velocities $v_{imp,fluid}^{(0)}$ may in principle be different. Results [1] suggest that they actually are equal, but it is not of concern here. Calculation of the marginal probability $P(v_{rel})$ shows that its asymptotic behaviour is, like Eq. (16),

$$P(v_{rel}) \sim \frac{1}{(v_{rel}^2)^{\sigma_{fluid}}} \qquad (22)$$

*irrespective* of the value $\sigma_{imp}$. The rest of the calculation then goes on just like in (17,18) and shows that the high-energy tail is completely determined by the background fluid.

## V. Conclusions

In this paper we have been able to provide a self-consistent derivation of NESM PDF's from microscopic dynamics, thereby providing a confirmation that Tsallis' is more than a useful but mere formalism. Rather, the appearance of non-standard statistics is a natural consequence of basic principles.

We point out that, even though the straightforward application of Beck's approach would be sufficient to recover the sought result, much more physical insight is gained by Sattin and Salasnich's approach.

Our results are not yet fully satisfactory. The most serious shortcoming, in our opinion, is that NESM formalism looks unable to provide self-consistently a quantitative estimate for its characteristic parameters (the exponent $\sigma$ in our case, $q$ in the general case). We cannot say if it is due to some deficiency of our own approach, which is not able to fully exploit all the informations available, or it is inherent to NESM framework.

*Acknowledgements*

The author wishes to thank Prof. Bettolo Marconi for suggesting him this theme of research, and L. Salasnich for useful discussions.

## Appendix A

After completion of this work, we became aware of the paper [15], where it is shown that Tsallis-type macroscopic distributions $P(x)$ can be obtained as stationary solutions of Fokker-Planck equations (or of equivalent Langevin equations)

$$\frac{d}{dt}P(x,t) = \frac{d}{dx}(K(x),P(x,t)) + \frac{1}{2}\frac{d^2}{dx^2}(D(x)P(x,t)) \qquad (A1)$$

provided that the convection and diffusion coefficients $K, D,$ are suitable function of the independent variable $x$: Borland gives general conditions that the couple of functions

*(K(x), D(x))* must satisfy, as well as showing several particular cases. Within this framework, Beck's ansatz (Eq. 10) is equivalent to making a particular choice for *K*, *D*, while Sattin and Salasnich's, instead, correspond to choosing arbitrary functions such that the resulting *P(x)* could even be not a Tsallis PDF. Note, however, that Borland's results do not cover all possible cases: for example, the *P(v)'s* studied in this work (Eqns. 3,4) satisfy master equations that do not reduce to the simple Fokker-Planck equation (A1) [1,2].

## Appendix B

To start with, we summarize the set of parameters used in the simulations [1]. They are: I) $\zeta \gg 1$ (very light impurities); II) $p = 1/2$ (equal fractions of impurities and fluid); III) $r_{\alpha\beta} \equiv r \sim 1 \forall \alpha, \beta$ (quasi-elastic restitution coefficients, independent of the species involved in the collision; the precise value reported in [1] was 0.95).

Conditions (I) and (III) allow to define the coefficients (Eqns. 3,4 of Ref. [4])

$$P = \frac{1-r}{2} \sim 0, Q = \frac{\frac{1}{\xi} - r}{1 + \frac{1}{\xi}} \sim -r \sim -1 \quad . \tag{B1}$$

We give now without proof some results from Ref. [4]: I) the coefficients $B^{(n)}$ appearing in Eq. (3) can be written as linear combinations of other coefficients $A_n$. Here we are interested only to $B^{(2)}$: its form is shown to be

$$B^{(2)} = 1 - 3A_2 + 3A_3 \qquad . \tag{B2}$$

The coefficients $A_n$ satisfy a recursion relation:

$$A_n = \frac{Q^{n-1}(1-Q) - P(1-P)}{1 - Q^n - nP(1-P)} A_{n-1} \qquad , \tag{B3}$$

where *P, Q* are defined in (B1), and which is supplemented by the initial condition $A_0 = 1$. It is straightforward to work out for our case

$$A_2 \sim \frac{-2}{1-Q^2} \rightarrow |A_2| \gg 1, \text{ and } A_3 \sim A_2 \tag{B4}$$

which, when replaced in (B2), give $B^{(2)} \sim 0$.

The set of parameters chosen in [1] is therefore such to almost cancel the coefficient of the leading term $\sim 1/(1 + v^2)^2$. Although this term will eventually dominate for large enough $v$, numerical simulations are unable to sample effectively that region, and instead are dominated for finite $v$ by the next-to-leading term $1/(1 + v^2)^3$, whose coefficient is much larger.